\documentclass[twocolumn]{aastex63}

\usepackage{savesym}
\savesymbol{tablenum}
\usepackage{siunitx}
\usepackage{bm}
\usepackage{mathtools}

\newcommand{\citeg}[1]{\citep[e.g.,][]{#1}}

\received{June 1, 2019}
\revised{January 10, 2019}
\accepted{\today}

\submitjournal{ApJ}

\shorttitle{Retrograde Orbiters in an AGN Disk}
\shortauthors{Secunda et al.}

\graphicspath{{./}{figures/}}

\begin{document}

\title{Evolution of Retrograde Orbiters in an AGN Disk}

\correspondingauthor{Amy Secunda}
\email{asecunda@princeton.edu}

\author{Amy Secunda}
\affil{Department of Astrophysics, American Museum of Natural History,
  Central Park West at 79th Street, New York, NY 10024, USA}
  \affil{Department of Astrophysical Sciences, Princeton University, Peyton Hall, Princeton, NJ 08544, USA}
 
\author{Betsy Hernandez}
\affil{Department of Astrophysics, American Museum of Natural History,
  Central Park West at 79th Street, New York, NY 10024, USA}
  \affil{Department of Astrophysical Sciences, Princeton University, Peyton Hall, Princeton, NJ 08544, USA}

\author{Jeremy Goodman}
  \affil{Department of Astrophysical Sciences, Princeton University, Peyton Hall, Princeton, NJ 08544, USA} 

\author{Nathan W. C. Leigh}
  \affiliation{Departamento de Astronom\'ia, Facultad de Ciencias F\'isicas y Matem\'aticas, Universidad de Concepci\'on, Concepci\'on, Chile}
\affiliation{Department of Astrophysics, American Museum of Natural History, Central Park West at 79th Street, New York, NY 10024, USA}

\author{Barry McKernan}
\affiliation{Department of Astrophysics, American Museum of Natural
  History, Central Park West at 79th Street, New York, NY 10024, USA}
\affiliation{Department of Science, Borough of Manhattan Community
  College, City University of New York, New York, NY 10007}
\affiliation{Physics Program, The Graduate Center, CUNY, New York, NY 10016}
\affiliation{Center for Computational Astrophysics, Flatiron Institute, New York, NY, 10010, USA}

\author{K.E. Saavik Ford}
\affiliation{Department of Astrophysics, American Museum of Natural
  History, Central Park West at 79th Street, New York, NY 10024, USA}
\affiliation{Department of Science, Borough of Manhattan Community
  College, City University of New York, New York, NY 10007}
\affiliation{Physics Program, The Graduate Center, CUNY, New York, NY 10016}
\affiliation{Center for Computational Astrophysics, Flatiron Institute, New York, NY, 10010, USA}

\author{Jose I. Adorno}
\affiliation{Department of Astrophysics, American Museum of Natural
  History, Central Park West at 79th Street, New York, NY 10024, USA}
\affiliation{NASA Goddard Space Flight Center, Greenbelt, MD 20771, USA}
\affiliation{Department of Physics, Queens College, City University of New York, Queens, NY, 11367, USA}

\begin{abstract}
AGN disks have been proposed as promising locations for the mergers of stellar mass black hole binaries (BBHs). Much recent work has been done on this merger channel, but the majority focuses on stellar mass black holes (BHs) orbiting in the prograde direction. Little work has been done to examine the impact of retrograde orbiters (ROs) on the formation and mergers of BBHs in AGN disks. Quantifying the retrograde contribution is important, since roughly half of all orbiters should initially be on retrograde orbits when the disk forms. We perform an analytic calculation of the evolution of ROs in an AGN disk. Because this evolution could cause the orbits of ROs to cross those of prograde BBHs, we derive the collision rate between a given RO and a given BBH orbiting in the prograde direction. In the examples given here, ROs in the inner region of the disk experience a rapid decrease in the semimajor axis of their orbits while also becoming highly eccentric in less than a million years. This rapid orbital evolution could lead to extreme mass ratio inspirals detectable by the Laser Interferometer Space Antenna. The collision rates of our example ROs with prograde BBHs in the migration trap depend strongly on the volume of the inner radiation-pressure-dominated region which depends on the mass of the supermassive black hole (SMBH). Rates are lowest for larger mass SMBHs, which dominate the AGN merger channel, suggesting that merger rates for this channel may not be significantly altered by ROs.
\end{abstract}

\keywords{black holes}

\section{Introduction} \label{sec:intro}

Active galactic nucleus (AGN) disks are promising locations \citep{mckernan2012,mckernan14,bellovary,bartos_2017a,stone_2017,mckernan18,leigh,secunda,secunda_2020,Yang_2019a,yang_2019b,Tagawa_2020,mckernan19, Gr_bner_2020,ishibashi_2020,LIGO_2020} for producing the stellar mass black hole binary (BBH) mergers detected by the Advanced Laser Interferometer Gravitational Wave Observatory (aLIGO) and Advanced Virgo \citep{Acernese_2014,aasi_2015,Abbott_2019}. An AGN disk is a favorable location for BBH mergers detectable by aLIGO because the gas disk will act to decrease the inclination of intersecting orbiters and harden existing BBHs  \citep{mckernan14,Yang_2019a}. Additionally, stellar mass black holes (BHs) on prograde orbits will exchange energy and angular momentum with the gas disk, causing migration in both the inward and outward radial directions \citep{bellovary,secunda,secunda_2020}. In particular these orbiters will migrate towards regions of the disk where positive and negative torques cancel out, known as migration traps. As these prograde orbiters (POs) migrate towards migration traps, they will encounter each other at small relative velocities.  Consequently, BBHs form that could merge on timescales of $10-500$~years \citep{baruteau2011,mckernan2012,mckernan18,leigh,baruteau_lin}. 

Despite an abundance of recent publications on BHs in AGN disks, thus far studies have largely ignored the impact of retrograde orbiters (ROs) in an AGN disk \citep[see however,][]{Sanchez_Salcedo_2020}. We could expect that since bulges have little net rotation, perhaps nuclei lack net rotation as well. Consequently, roughly half of the initial BH population of a nuclear star cluster should be on retrograde orbits when the gas disk forms. While some initially inclined orbiters will flip from retrograde to prograde orbits as they are ground down into alignment with the disk \citep[][]{rauch_1995,MacLeod_2020}, the population of orbiters initially aligned with the disk or on slightly inclined orbits should be roughly half retrograde. These ROs will be impacted by the disk in a significantly different way from POs due to their larger velocities relative to the gas disk. Additionally, ROs will encounter POs in the disk with large relative velocities, meaning they are less likely to form BBHs with POs and more likely to ionize binaries in the disk \citep{leigh}. Therefore ROs could have a significant affect on the number of BBHs and mergers in AGN disks.

 We aim to calculate the evolution of BHs initially orbiting in the retrograde direction when the gas disk appears, and predict whether these ROs interact with POs. In \S \ref{sec:evolve} we derive eqs. for the time evolution of the semimajor axis and eccentricity of a RO in an AGN disk. In \S \ref{sec:prob} we derive the collision rate between a RO and a BBH orbiting in the prograde direction as a function of their semimajor axes and eccentricities. We use these derivations to give three fiducial examples at two fiducial SMBH masses in \S \ref{sec:results}. Finally, in \S \ref{sec:discuss} we discuss the implications of the results of our model for gravitational and electromagnetic wave detections of BHs and BBHs in AGN disks.

\section{Orbital Evolution}
\label{sec:evolve}
Here we derive eqs. for the time evolution of the semimajor axis, $a$, and eccentricity, $e$, of a RO in an AGN disk. We assume that the disk is Keplerian and axisymmetric, and that the RO has already settled into an orbit on the disk midplane.

For a BH orbiting in an AGN disk in the retrograde direction the relative velocity ($\textbf{\textit{v}}_{\rm{rel}}=\textbf{\textit{v}}-\textbf{\textit{v}}_{\rm{disk}}$) between the orbiter and the disk is highly supersonic, with Mach number $v_{\rm{rel}}/c_{\rm{s}} \sim (h/r)^{-1} \gg 1$, where $h/r$ is the disk aspect ratio, and $c_{\rm{s}}$ is the isothermal sound speed at the midplane. The gas drag force on a BH of mass $m$ can be approximated as dynamical friction \citep{1987_binney,1999_ostriker},
\begin{equation}
    \label{eq:fdrag}
    \textbf{F}_{\rm{drag}} = -\frac{4\pi \ln{\Lambda}(Gm)^2 \rho}{v_{\rm{rel}}^3} \bm{v}_{\rm{rel}},
\end{equation}
where $\rho$ is the local mass density of the disk, and $\Lambda \sim hv_{\rm{rel}}^2/Gm$, where $h$ is the scale height of the disk and $G$ is the gravitational constant. We assume $m$ is small enough that $\Lambda \gg 1$. The additional contribution to the drag from Bondi-Hoyle-Lyttleton accretion onto the BH will be smaller by a factor of $\sim (\ln{\Lambda})^{-1}$, and so we neglect it. 

The orbital energy of the BH is,
\begin{equation}
    \label{eq:energy}
    E = -\frac{GMm}{2a},
\end{equation}
where $M$ is the mass of the SMBH. Because we neglect accretion onto the BH,
\begin{equation}
\label{eq:lna_lnE}
    \frac{d \ln{a}}{dt} = -\frac{d\ln{E}}{dt},
\end{equation}    
and
\begin{equation}
\label{eq:dedt}
    \frac{dE}{dt}=\textbf{F}_{\rm{drag}} \cdot \bm{v} = -\frac{4\pi \ln{\Lambda}(Gm)^2 \rho}{v_{\rm{rel}}^3} \bm{v} \cdot \bm{v}_{\rm{rel}},
\end{equation}
where $\bm{v}$ is the velocity of the RO. Defining the angular momentum of the AGN disk as positive, the angular momentum for a RO becomes
\begin{equation}
    \label{eq:ang_mom}
    L = -m\sqrt{GMa(1-e^2)}.
\end{equation}
The torque on the orbiter is
\begin{equation}
\label{eq:dldt}
    \frac{dL}{dt} = \bm{F}_{\rm{drag}} \cdot \bm{\hat{e}}_{\rm{\phi}}r = -\frac{4\pi \ln{\Lambda}(Gm)^2 \rho}{v_{\rm{rel}}^3} (rv_{\rm{\phi}} - \sqrt{GMr}),
\end{equation}
where $\bm{\hat{e}}_{\rm{\phi}}$ is a unit vector in the azimuthal direction, and $r$ is the radial distance of the RO from the central SMBH.

Using eqs. \ref{eq:energy} and \ref{eq:ang_mom}, for small changes $da$, $de^2$ in $a$ and $e^2$ we get
\begin{equation}
\label{eq:dE}
    dE = \frac{GMm}{2a^2}da,
\end{equation}
and
\begin{equation}
\label{eq:dL}
    dL = -\frac{m}{2}\left[\sqrt{\frac{GM}{a}(1-e^2)}da - \sqrt{\frac{GMa}{1-e^2}}de^2\right].
\end{equation}
Using eq.~\eqref{eq:dE} to put eq.~\eqref{eq:dL} in terms of $dE$ instead of $da$ and substituting in the mean motion, $n = -\sqrt{GM/a^3}$, gives the change in eccentricity in terms of the change in energy and angular momentum,
\begin{equation}
    \frac{de^2}{dt} = \frac{2a}{GMm}(1-e^2)\left(\frac{dE}{dt} - \frac{n}{\sqrt{1-e^2}}\frac{dL}{dt}\right).
\end{equation}

Using the fact that $E = m(v^2/2 - GM/r)$, where $v$ is the magnitude of the velocity, $\bm{v}$, $L=mrv_{\rm{\phi}}=m\sqrt{GMa(1-e^2)}$ and $\bm{v}_{\rm{disk}} = \bm{\hat{e}}_{\rm{\phi}} \sqrt{GM/r}$, we obtain
\begin{equation}
\label{eq:vminvd}
    |\bm{v}-\bm{v}_{\rm{disk}}|^2 = -\frac{GM}{a} + 3\frac{GM}{r} + 2GM\sqrt{a(1-e^2)}r^{-3/2},
\end{equation}
\begin{equation}
    \label{eq:vdot_vmvd}
    \bm{v} \cdot (\bm{v} - \bm{v}_{\rm{disk}}) = -\frac{GM}{a} + 2\frac{GM}{r} + GM\sqrt{a(1-e^2)}r^{-3/2},
\end{equation}
which allows us to eliminate the velocities in eqs. \ref{eq:dedt} and \ref{eq:dldt} in favor of $r$. $r$, as a function of the azimuthal angle $\phi$, is
\begin{equation}
    \label{eq:r_of_phi}
    r = \frac{a(1-e^2)}{1+e\cos{(\phi - \phi_{\rm{p}})}},
\end{equation}
where $\phi_{\rm{p}}$ is the angle at pericenter.  We set $\phi_{\rm{p}} =$ 0, because our disk is axisymmetric. By Kepler's Second Law, the time interval $dt$ corresponding to the angular interval $d\phi$ are related by,
\begin{equation}
    \label{eq:dt_overP}
    \frac{dt}{P} = \frac{r^2d\phi}{2\pi a^2\sqrt{1-e^2}},
\end{equation}
where $P$ is the orbital period. We can use eq.~\eqref{eq:dt_overP} to write the average change in energy, angular momentum, and eccentricity over one orbital in terms of $d\phi$. 

Apart from the velocities, $dE/dt$ and $dL/dt$ depend on $r$ through the midplane density $\rho(r)$ and $\Lambda$. Here we take $\rho(r)\propto r^{\gamma}$ and ignore the slight variation in $\ln{\Lambda}$ along the orbit.

We define
\begin{equation}
    \label{eq:f_of_a}
    f(a) = 4\pi \ln{\Lambda}(Gm)^2 \rho(a) \sqrt{a/GM},
\end{equation}
which has the same dimensions as $dE/dt$ and $ndL/dt$. We can now write
\begin{equation}
    \label{eq:avg_dedt}
    \langle \frac{dE}{dt} \rangle = -f(a) I_{\rm{E}}(\gamma,e),
\end{equation}
\begin{equation}
    \label{eq:avg_dLdt}
    \langle \frac{dL}{dt} \rangle = -\frac{f(a)}{n}I_{\rm{L}}(\gamma,e),
\end{equation}
\begin{equation}
    \label{eq:avg_dadt}
    \langle \frac{da}{dt} \rangle = -f(a)\frac{2a^2}{GMm} I_E(\gamma,e),
\end{equation}
and
\begin{equation}
    \label{eq:avg_deccdt}
    \langle \frac{de^2}{dt} \rangle = -f(a) \frac{2a}{GMm}(1-e^2) [I_{\rm{E}}(\gamma,e) + \frac{1}{\sqrt{1-e^2}}I_{\rm{L}}(\gamma,e)],
\end{equation}
where $I_{\rm{E}}$ and $I_{\rm{L}}$ are the dimensionless integrals,
\begin{equation}
    \label{eq:I_E}
    I_{\rm{E}}(\gamma,e) = \frac{1}{2\pi \sqrt{1-e^2}}\int_0^{2\pi} \frac{(-1+2u+\sqrt{1-e^2}u^{3/2})u^{-\gamma-2} d\phi}{(-1+3u+2\sqrt{1-e^2}u^{3/2})^{3/2}} 
\end{equation}
and
\begin{equation}
    \label{eq:I_L}
    I_{\rm{L}}(\gamma,e) = \frac{1}{2\pi \sqrt{1-e^2}}\int_0^{2\pi} \frac{(-\sqrt{1-e^2}-u^{-1/2})u^{-\gamma - 2} d\phi}{(-1+3u+2\sqrt{1-e^2}u^{3/2})^{3/2}},
\end{equation}
where,
\begin{equation}
    \label{eq:u_of_phi}
    u(\phi) \equiv a/r = \frac{1+e\cos{\phi}}{1-e^2}.
\end{equation}
These eqs. can be integrated numerically to solve for the eccentricity and semimajor axis of a RO as a function of time for a given set of disk parameters. We discuss the orbital evolution of ROs with three different initial eccentricities in \citet{Sirko:2003aa} AGN disks with two different SMBH masses in \S \ref{sec:results}.

\section{Collision Rates}
\label{sec:prob}

In this section we briefly outline a derivation of the collision rate of a RO (body 1) and a prograde BBH (body 2). For additional detail see Appendix \ref{sec:appendix}. We assume that the apsidal precession rate due to both relativistic effects and disk self-gravity is rapid compared to the interaction rate, such that the probability of finding an orbiter in a given area element $rdrd\phi$ is independent of azimuth, $\phi$. Therefore, the collision probability is proportional to the fraction $(dt/P)_i$ of the orbit of body $i$ spent between $r$ and $r+dr$,
\begin{align}
  \label{eq:annulus}
  \frac{d\mathbb{P}_i}{dr}dr &= \frac{2}{P_i}\frac{dr}{|v_r|} \ = \frac{2}{P_i}\frac{dr}{\sqrt{2[\hat{E}-\hat{\Phi}(r)-\hat{L}^2/2r^2]}}\nonumber\\[1ex]
  &= \frac{1}{\pi a_i} \frac{ r dr}{\sqrt{(r_{+,i}-r)(r-r_{-,i})}}\,,\qquad r_{\pm,i} =  a_i(1\pm e_i)\,
\end{align}
where $\hat{E}$, $\hat{\Phi}$, and $\hat{L}$, are the total energy, potential energy, and angular momentum per unit mass. The factor of 2 occurs in the numerator because the orbit crosses a given radius $r$ twice per orbit, provided that $a(1-e)<r<a(1+e)$. The second line follows from the relations $P= 2\pi\sqrt{a^3/GM}$, $\hat{E}=-GM/2a$, $\hat{\Phi}(r)=-GM/r$ and $\hat{L}^2=GMa(1-e^2)$.

Orbiters in an AGN disk will be excited onto slightly inclined orbits by turbulent motions in the disk, but their inclination will also be damped by drag forces from the gas. Without a specific model for turbulence, we assume for simplicity that the probability of finding an orbiter at height $z$ to $z+dz$ is gaussian,
\begin{equation}
  \label{eq:hintro}
  \frac{d\mathbb{P}_i}{dz}dz=\frac{\exp(-z^2/2h_{\rm{BH}}^2)}{\sqrt{2\pi h_{\rm{BH}}^2}}\,dz\,,
\end{equation}
with a scale height $h_{\rm{BH}}$ estimated as (see Appendix \ref{sec:appendix} for details),
\begin{equation}
    \label{eq:h_bh}
    h_{\rm{BH}} \simeq h \alpha^{1/2} \left(\frac{\alpha^{3/2} \rho h^3}{m_1}\right)^{1/2},
\end{equation}
where $\alpha$ is the Shakura–Sunyaev \citep{shakura_sunyaev} viscosity parameter. For simplicity we assume this scale height is constant.

Since the area of the annulus is $2\pi r dr$ and the distribution over height is given by
eq.~\eqref{eq:hintro}, the probability per unit volume $dV=rdrd\phi dz$ of finding the body
near a given point $(r,\phi,z)$ is
\begin{multline}
  \label{eq:dPdV}
  \frac{d\mathbb{P}_i}{dV} =\frac{1}{r} \frac{d\mathbb{P}_i}{dr}\frac{d\mathbb{P}_i}{d\phi}\frac{d\mathbb{P}_i}{dz}
  \\
  =\frac{1}{2\pi^2a_i(2\pi h_{\rm BH}^2)^{1/2}}\frac{\exp(-z^2/2h_{\rm BH}^2)}{\sqrt{(r_{+,i}-r)(r-r_{-,i})}}.
\end{multline}
We will assume $e_2\sim$~0, because the disk acts to circularize POs \citep{tanaka_ward}. This assumption gives
\begin{equation}
    \label{eq:prob2}
  \frac{d\mathbb{P}_2}{dV} \ \approx\  \frac{\exp(-z^2/2h_{\rm BH}^2)}{\sqrt{2\pi h_{\rm BH}^2}}\ \frac{\delta(r-a_2)}{2\pi^2 a_2}\,,
\qquad\mbox{if $e_2\ll 1$,}
\end{equation}
where $\delta(r-a_2)$ is a delta function centered at $a_2$.

If the annuli of the two bodies overlap, the expected interaction rate between them becomes
\begin{align}
  \label{eq:rate}
  \tau_{\rm coll}^{-1} = \int dV \frac{d\mathbb{P}_1}{dV}\frac{d\mathbb{P}_2}{dV}\,v_{\rm 12}\sigma(v_{\rm 12}).
\end{align}

If we assume the $z$-components of the velocities of the orbiters are negligible and that $e_2 \sim 0$, as above, the relative velocity between the RO and the BBH is
\begin{equation}
    \label{eq:vrel}
    v_{12}=\left[GM\left(\frac{3}{a_2} -\frac{1}{a_1}
    +\frac{2}{a_2^{3/2}}\sqrt{a_1(1-e_1^2)}\right)\right]^{1/2},
\end{equation}
(see Appendix \ref{sec:appendix} for a detailed derivation).

The interaction cross section of the BBH and the RO is,
\begin{equation}
    \label{eq:sigma}
    \sigma \sim \pi s_{\rm bin}^2 f^2 \ln(1/f).
\end{equation}
 Here $s_{\rm bin}$ is the semimajor axis of the binary itself, which we take to be the mutual Hill radius of the two BHs in the binary (with total mass $m_2$),
 \begin{equation}
	\label{eq:rhill}
	R_{\rm mH} = r_{\rm 2}\left(\frac{m_{\rm 2}}{3M}\right)^{1/3}.
\end{equation} 
$f$ is the Safronov number, a dimensionless ``gravitational focusing" factor,
\begin{equation}
    \label{eq:focusing}
    f \equiv \frac{Gm_2}{s_{\rm bin}v_{12}^2}.
\end{equation}
$\sigma$ is taken in the limit that $f \ll 1$. Because $v_{12}$ will be very large, the encounters will be fast and gravitational focusing will not be important.

Integrating eq.~\eqref{eq:rate} gives,
\begin{multline}
\label{eq:simprate}
  \tau_{\rm coll}^{-1} =
  \frac{1}{\sqrt{4\pi h_{\rm BH}^2}}\,\times\,
  \\
  \frac{1}{2\pi^3 a_1\sqrt{(a_1(1+e_1)-a_2)(a_2-a_1(1-e_1))}}\,\times\, v_{12}\sigma(v_{12})\,,
\end{multline}
with $v_{12}$ given by eq.~\eqref{eq:vrel} and $\sigma(v_{12})$ by eq.~\eqref{eq:sigma} (see Appendix \ref{sec:appendix} for more details). The two terms in parentheses in the denominator of the second term define the limits where a collision is possible given our assumptions, since both terms must be positive. That is, it is not possible for a collision to take place if $a_2$ is greater than the apocenter of the RO or less than the pericenter of the RO. We discuss the collision rate for three different fiducial $e_1$ in \S \ref{sec:results}.

\begin{figure*}
    \centering
    \begin{tabular}{lr}
     \includegraphics[width=0.48\textwidth]{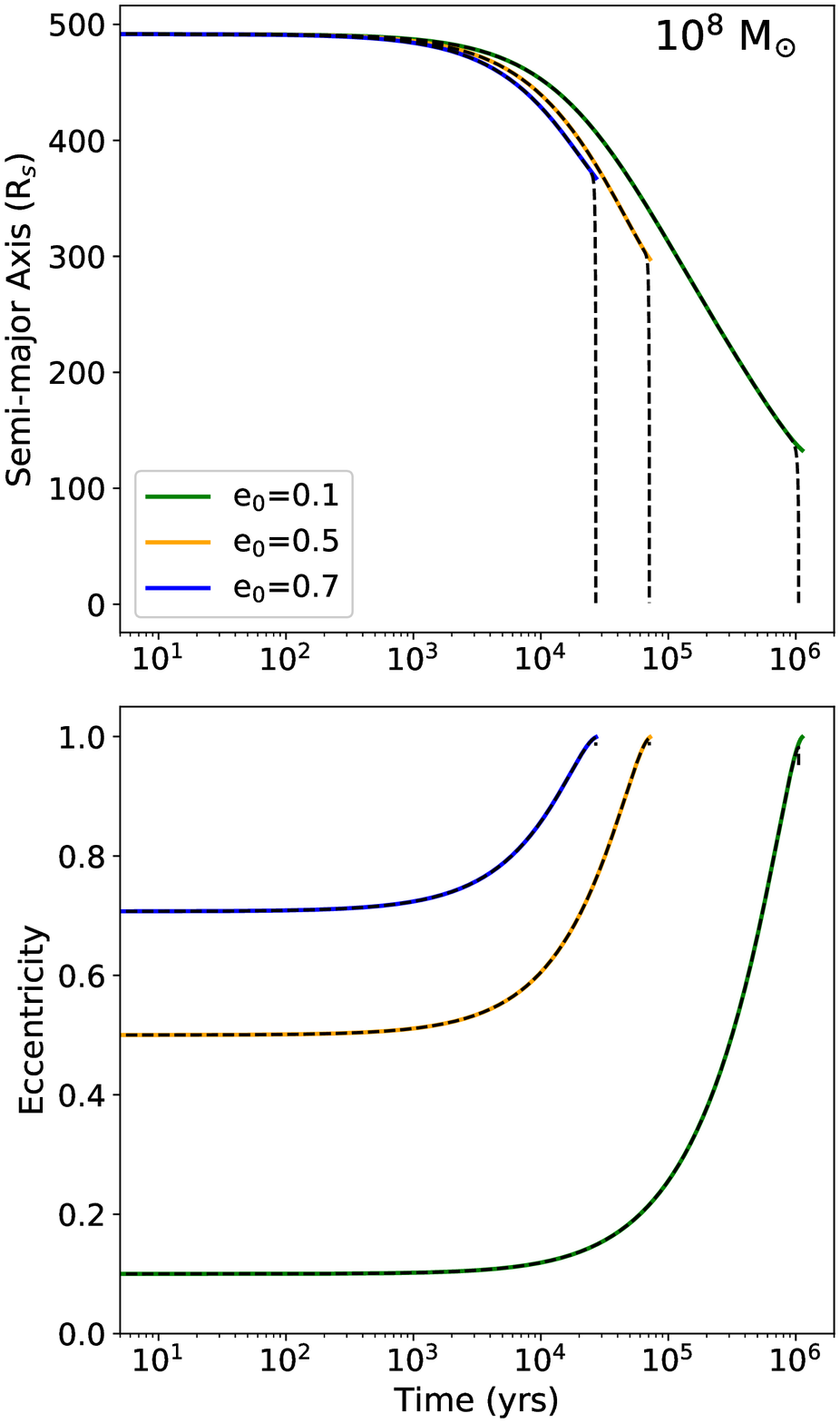}
    & \includegraphics[width=0.48\textwidth]{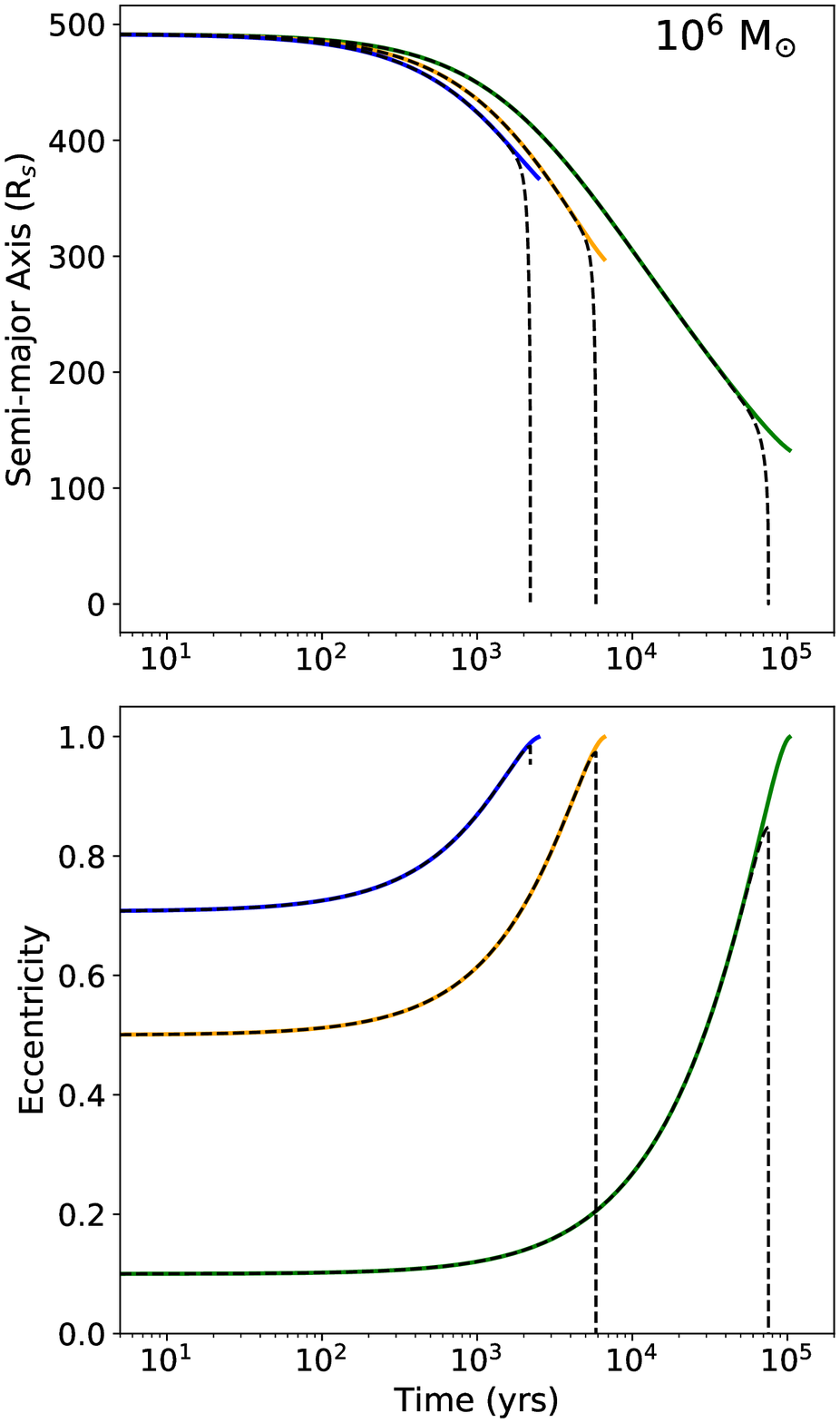}
    \end{tabular}
    \caption{The colored lines show the evolution of the semimajor axis (top panel) and eccentricity (bottom panel) of 10~M$_{\rm{\sun}}$ ROs with different initial eccentricities, calculated by numerically integrating eqs. \ref{eq:avg_dedt}, \ref{eq:avg_dLdt} \ref{eq:avg_dadt}, and \ref{eq:avg_deccdt} in \S \ref{sec:evolve}. The dashed black lines show the evolution of these orbiters when we include GW circularization \citep{peters} in our integration. All orbiters begin with a semimajor axis of 500~R$_{\rm{s}}$ and are evolved until they reach an eccentricity of 0.999 for the integration that does not include GW circularization, or merge with the SMBH for the integration that does. The left and right panels show these values for a RO in a \citet{Sirko:2003aa} disk with a $10^8$~M$_{\sun}$ and $10^6$~M$_{\sun}$ SMBH, respectively.}
    \label{fig:ret_evolve}
\end{figure*}

\section{Results}
\label{sec:results}
We choose to test our formalism on a \citet{Sirko:2003aa} AGN disk with two different SMBH masses, 10$^8$~M$_{\sun}$ and 10$^6$~M$_{\sun}$. The first mass is chosen because \citet{Tagawa_2020} found that the AGN channel for LIGO mergers is dominated by disks surrounding 10$^{7-8}$~M$_{\sun}$ SMBHs. We choose the second mass because the Laser Interferometer Space Antenna (LISA) is most sensitive to extreme mass ratio inspirals (EMRIs) where the SMBH is $10^5 - 10^6$~M$_{\rm{\sun}}$ \citep{Babak_2017}. Preliminary results suggest that SMBH mass is a more influential parameter than the specific AGN disk model chosen.

The \citet{Sirko:2003aa} AGN disk model is a modification of the \citet{shakura_sunyaev} Keplerian viscous disk model with a constant high accretion rate fixed at Eddington ratio 0.5. The disk is marginally stable to gravitational fragmentation, although an additional unknown heating mechanism is assumed to maintain stability in the outer disk. Due to the unknown physics of the outer disk, here we focus on the inner disk which is stable ($Q>1$) without invoking this additional heating mechanism. We also choose to focus on the inner disk region, because that is where the migration trap is located and where BHs must pass through in order to become EMRIs. In this region of the disk, mass density increases with radius and radiation pressure is dominant. We refer to this region as the inner radiation-pressure-dominated region, because while other AGN disk models treat heating differently, such as \citet{thompson}, these models still have a transition from gas to radiation pressure domination in the inner disk.

The distance from the SMBH at which a \citet{Sirko:2003aa} disk transitions from radiation to gas pressure dominant is roughly proportional to the mass of the SMBH, because the ratio of gas pressure to total pressure depends on the distance to the SMBH in units of the Schwarzschild radius, $R_{\rm{s}}$, with only a weak additional dependence on SMBH mass \citep[see eq. A3 in][]{goodman_2003}. In addition, although the aspect ratio of the disk ($h/r$) is only weakly dependent on disk mass, because a larger SMBH mass increases the radius of the radiation-pressure-dominated region, the scale height ($h$) will increase as SMBH mass increases. Therefore the volume of the inner radiation-pressure-dominated region will be significantly larger for a larger mass SMBH, which has implications for the collision rates of objects within this region (see \S \ref{sec:discuss}).

\begin{figure}
\begin{tabular}{ll}
\centering
     \includegraphics[width=0.48\textwidth]{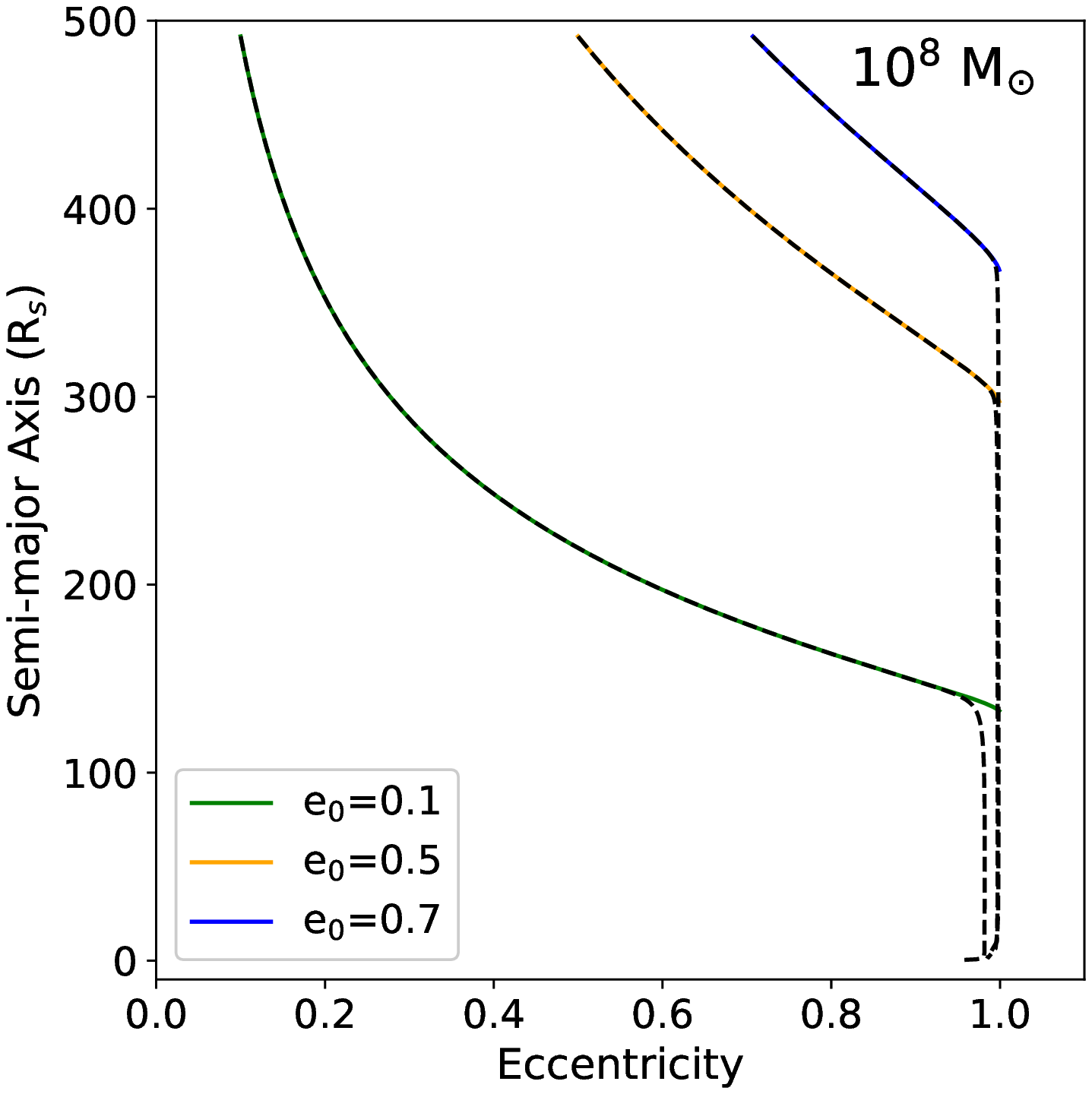}
    & \\
    \includegraphics[width=0.48\textwidth]{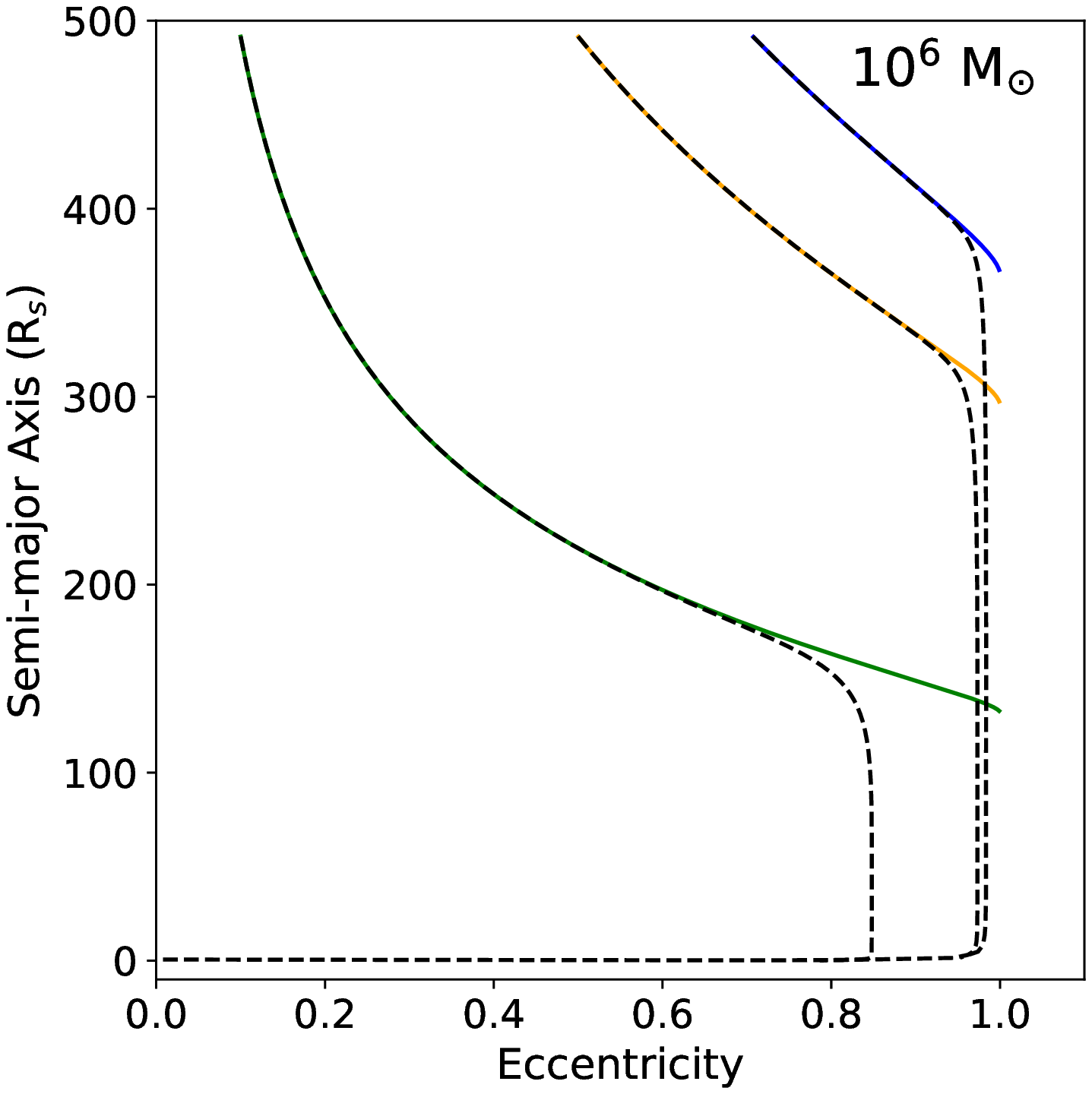}
    \end{tabular}
    \caption{The evolution of the semimajor axis and eccentricity of a RO in a \citet{Sirko:2003aa} AGN disk with a $10^8$~M$_{\sun}$ SMBH (top panel) and a $10^6$~M$_{\sun}$ SMBH (bottom panel) for the three example ROs in Figure \ref{fig:ret_evolve}. ROs evolve over time from the top left of the figure down.}
    \label{fig:eva}
\end{figure}

The solid lines in Figure \ref{fig:ret_evolve} show the evolution of the semimajor axes (top) and eccentricities (bottom) of 10~M$_{\sun}$ ROs orbiting a 10$^8$~M$_{\sun}$ (left panel) and a 10$^6$~M$_{\sun}$ (right panel) SMBH with initial eccentricities $e_{\rm{0}}$=0.1, 0.5, and 0.7, calculated through numerical integration of eqs. \ref{eq:avg_dedt}, \ref{eq:avg_dLdt}, \ref{eq:avg_dadt}, and \ref{eq:avg_deccdt}. All orbiters were initiated with  $a=500$~R$_{\rm{s}}$ and integrated over time until they reached $e=0.999$. Figure \ref{fig:eva} shows the relation between semimajor axis and eccentricity for the same example ROs with the same notation. The evolution in semimajor axis for the special case of a retrograde orbiter on a circular orbit is given in Appendix \ref{sec:circular}.

ROs at all $e_{\rm{0}}$ see a significant increase in their eccentricity, which is at least doubled, and decrease in their semimajor axis within $10^4$ and $10^5$~years for the 10$^6$~M$_{\sun}$ and 10$^8$~M$_{\sun}$ case, respectively. All orbiters reach an eccentricity of 0.999 in under $10^5$ or $10^6$~years, for the smaller and larger SMBH mass, respectively. ROs with greater $e_{\rm{0}}$ become highly eccentric on shorter timescales. For example, orbiters around a 10$^8$~M$_{\sun}$ SMBH with $e_{\rm{0}}\gtrsim$ 0.5 reach $e=0.999$ within 100~kyr. 

The dashed black lines in Figure \ref{fig:ret_evolve} show the evolution of ROs with the same initial conditions as the colored lines, when accounting for gravitational wave (GW) circularization \citep{peters}. We evolve $e$ and $a$ by the rates in \citet{peters} at the values we find for $e$ and $a$ after evolving them with eqs. \ref{eq:avg_dedt}, \ref{eq:avg_dLdt}, \ref{eq:avg_dadt} and \ref{eq:avg_deccdt} from \S \ref{sec:evolve}. These rates are integrated over time until $a=0$. GW circularization becomes more rapid as the eccentricity of the orbiter increases, slowing the eccentricity driving once a high eccentricity is reached. For the 10$^8$~M$_{\sun}$ SMBH, the maximum eccentricity reached is now 0.982, 0.997, and 0.998, and the eccentricity at the time of merger is 0.932, 0.984, and 0.983 for ROs with $e_{\rm{0}}$=0.1, 0.5, and 0.7, respectively. For the 10$^6$~M$_{\sun}$ SMBH, the maximum eccentricity reached is 0.848, 0.973, and 0.984 for ROs with $e_{\rm{0}}$=0.1, 0.5, and 0.7, respectively. The eccentricity at the time of merger is 0.953 for the RO with $e_{\rm{0}}$=0.7. For lower $e_0$, however, ROs have time to circularize before they merge.

\begin{figure}
\begin{tabular}{ll}
\centering
     \includegraphics[width=0.48\textwidth]{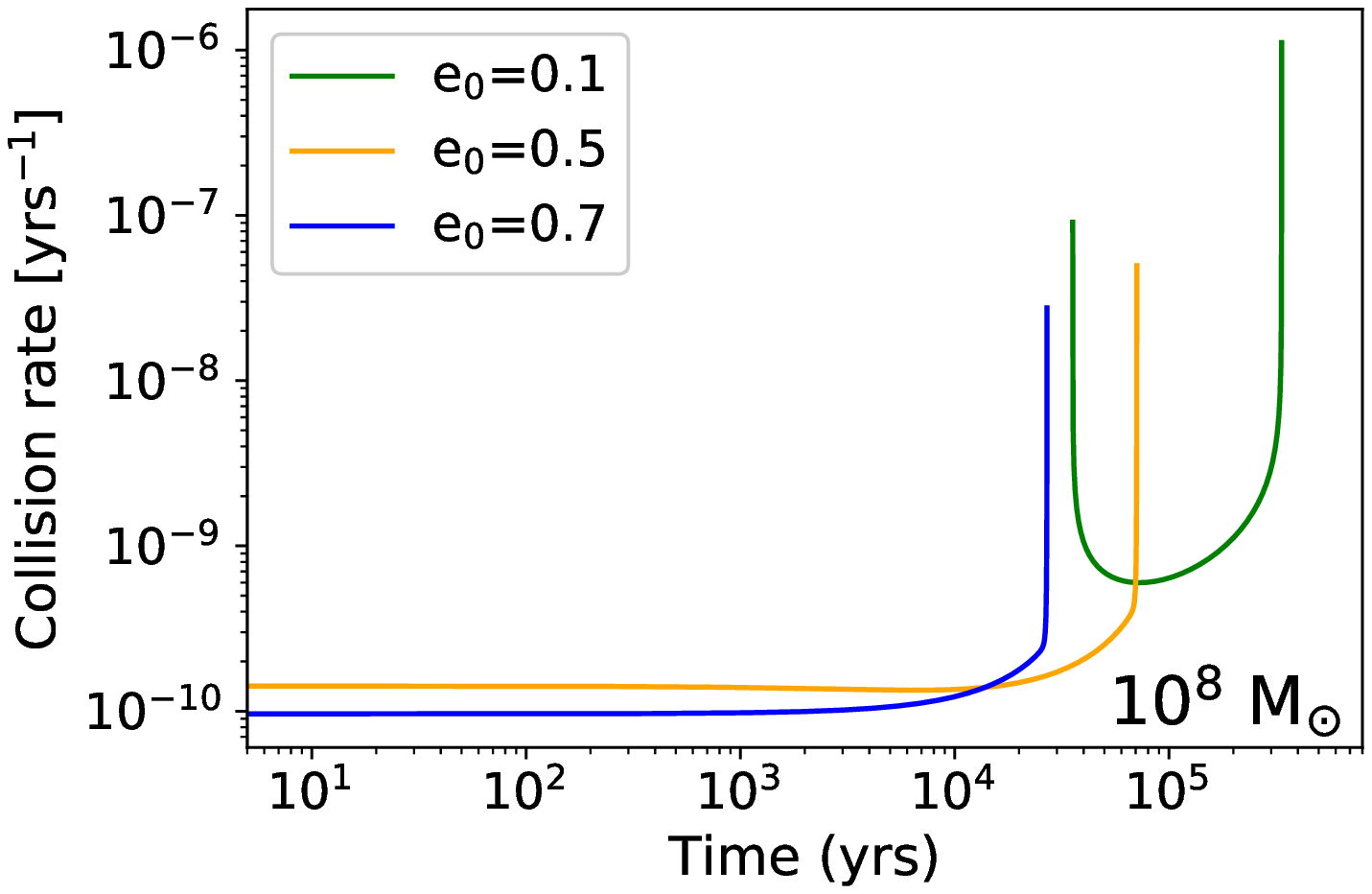}
    & \\
    \includegraphics[width=0.48\textwidth]{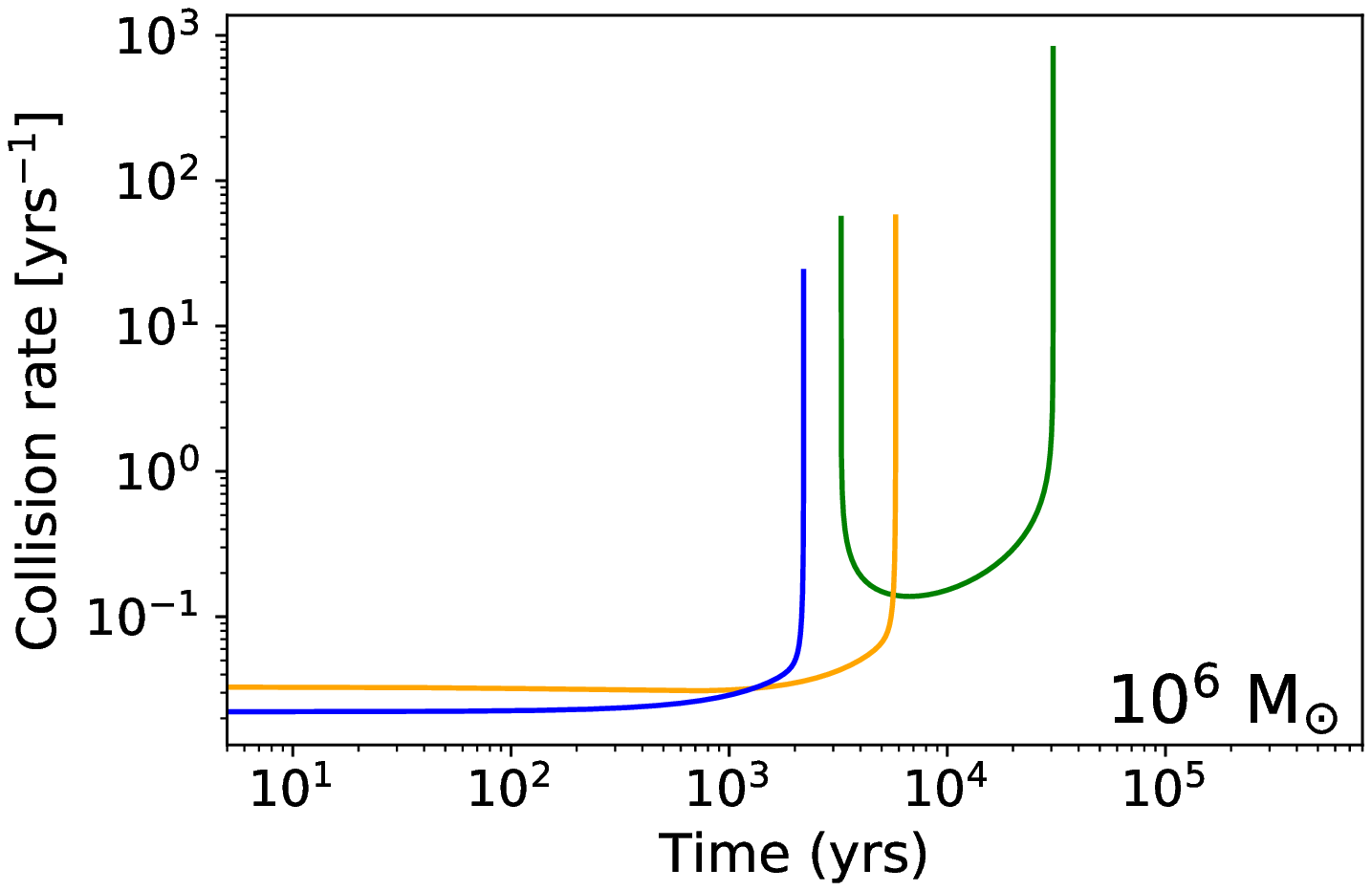}
    \end{tabular}
    \caption{The collision rate as a function of time predicted by eq.~\eqref{eq:simprate} for a RO with three different initial eccentricities, and a BBH orbiting in the prograde direction with respect to the disk on a circular orbit. The top and bottom panels show these values for orbiters in a \citet{Sirko:2003aa} disk with a $10^8$~M$_{\sun}$ and $10^6$~M$_{\sun}$ SMBH, respectively. In the top panel the integrated probability of an interaction occurring before the RO reaches a=0 is \num{5.0e-4}, \num{1.6e-5}, and \num{4.1e-6} for a RO with $e_{\rm{0}}$= 0.1, 0.5, and 0.7, respectively. In the bottom panel the integrated probability is greater than 1 for all three $e_{\rm{0}}$.}
    \label{fig:coll_rate}
\end{figure}

Figure \ref{fig:coll_rate} shows the collision rate of each orbit, $\tau_{\rm coll}^{-1}$, as a function of time calculated with eq.~\eqref{eq:simprate} of ROs around a $10^8$~M$_{\sun}$ and $10^6$~M$_{\sun}$ SMBH in the top and bottom panels, respectively, with the three example $e_{\rm{0}}$. The mass of the RO is 10~M$_{\sun}$ and the total mass of the binary $m_2=20$~M$_{\sun}$. For simplicity we take the semimajor axis of the BBH, $a_2$, to be constant at 330~$R_{\rm{s}}$, roughly the location of the migration trap in a \citet{Sirko:2003aa} AGN disk for both SMBH masses \citep{bellovary}. For these parameters $s_{\rm bin}=0.13,2.7$~AU for disks with $10^6$, $10^8$~M$_{\sun}$ SMBHs, respectively. $a_1$ and $e_1$ evolve over time as calculated above, with GW circularization and $a_1$=500~$R_{\rm{s}}$, initially. If $a_2$ is greater than the apocenter distance or less than the pericenter distance of the RO for a given orbit we take $\tau_{\rm coll}^{-1} = 0$.

We choose the initial parameters for this example to resemble the most common conditions in \citet{secunda} and \citet{secunda_2020}, who find that BHs migrate towards the migration trap at $\sim$~330~$R_{\rm{s}}$ in a \citet{Sirko:2003aa} AGN disk, where they start forming BBHs on timescales similar to the orbital evolution of ROs ($\sim 10^4-10^5$ years) and remain for the lifetime of the disk. This over-dense population of BBHs is a prime target for a RO to interact with. Preliminary tests show that changing the location of the BBH and having BBHs migrate within the inner radiation-pressure-dominated region of the disk does not have a significant affect on $\tau_{\rm coll}^{-1}$.

For ROs orbiting a $10^8$~M$_{\sun}$ SMBH with $e_{\rm{0}}$=0.5,0.7, $\tau_{\rm coll}^{-1}\sim\mathcal{O}(10^{-10})$~yr$^{-1}$ initially and increases to $\sim\mathcal{O}(10^{-8})$~yr$^{-1}$ as GW circularization becomes important. At first, $\tau_{\rm coll}^{-1}=0$ for the RO with $e_{\rm{0}}$=0.1, since its orbit will not cross the orbit of the BBH until its semimajor axis has decreased and its eccentricity has increased. Once the orbits do cross $\tau_{\rm coll}^{-1}$ starts out relatively high, around $\mathcal{O}(10^{-7})$~yr$^{-1}$. Then, while the eccentricity is still low, $\tau_{\rm coll}^{-1}$ decreases as the semimajor axis decreases reaching a minimum of around \num{6e-10}~yr$^{-1}$. Next, as the eccentricity increases, the decrease in semimajor axis causes $\tau_{\rm coll}^{-1}$ to increase to $\sim10^{-6}$~yr$^{-1}$. Finally, the semimajor axis becomes too small for the RO to cross the orbit of the BBH, and $\tau_{\rm coll}^{-1}=0$. 

The total probability of an encounter summed over all orbits before the RO reaches $a=0$ is \num{5.0e-4}, \num{1.6e-5}, and \num{4.1e-6} for $e_{\rm{0}}$=0.1, 0.5, and 0.7, respectively. Our calculations suggest that an interaction between a RO and a BBH orbiting in the prograde direction is most likely to occur for ROs with smaller $e_{\rm{0}}$, because their orbits have more time to evolve to smaller semimajor axes before they are driven to high eccentricities. However, our calculated probability of interaction is still very small for these orbiters, suggesting that the likelihood of an interaction between a prograde BBH and a RO is small for an SMBH this massive.

$\tau_{\rm coll}^{-1}$ is much larger for the $10^6$~M$_{\sun}$ SMBH case, ranging from $\sim$0.01--100~yr$^{-1}$ over time depending on $e_0$. These high rates suggest that ROs in a disk with a lower mass SMBH could collide with prograde BBHs multiple times over the course of their evolution.

\section{Discussion}
\label{sec:discuss}
ROs in the inner radiation-pressure-dominated regions of \citet{Sirko:2003aa} AGN disks with $10^8$~M$_{\sun}$ and $10^6$~M$_{\sun}$ SMBHs migrate inward on timescales of $10^4-10^5$ and $10^3-10^4$ years, respectively, depending on their initial eccentricity. They also experience a rapid increase in their eccentricity, reaching $e\gtrsim0.999$ or $e\gtrsim0.85$ in less than a megayear, without and with GW circularization, respectively. This eccentricity driving is a result of the instantaneous drag pulling the orbiter towards co-rotation with the local disk (see eq.~\ref{eq:fdrag}). Eq.~\ref{eq:dldt} shows this pull always results in a positive torque, because $v_\phi<0$. Since the angular momentum of the retrograde orbiter is negative, the positive torque decreases its absolute magnitude. Because the drag is strongest near apocenter, where $v_{\rm rel}$ is smallest and (in the inner radiation-dominated disk) the surface density is largest, the energy of the RO is less affected than its angular momentum, making the orbit more eccentric. The PO case is more complex, with co-rotation torques, which do not act on ROs, playing a large role in dampening the eccentricity of POs \citep{artymowicz_1993}. The timescale for the eccentricity dampening of a 10~M$_{\sun}$ PO given by \cite{tanaka_ward} would be on the order of a few hundred years for a $10^8$~M$_{\sun}$ SMBH and a few years for a $10^6$~M$_{\sun}$ SMBH, significantly shorter than the eccentricity driving timescale for ROs found here.

We have assumed ROs have already settled into the inner radiation-pressure-dominated region of the disk, since that is where the migration trap is located and where ROs will migrate through in order to become EMRIs. Preliminary results show that further out in the disk ROs will still experience a decrease in semimajor axis, but will have their eccentricities decreased, because the density of the disk decreases with radius at these radii. However, once their semimajor axes decrease sufficiently for ROs to reach the inner radiation-pressure-dominated disk the eccentricity driving shown here will begin. Therefore ROs initially in the outer disk will likely also become EMRIs.

GW circularization only has a minimal affect on our 10~M$_{\rm{\sun}}$ ROs until they reach $e\gtrsim0.8$. Once a maximum eccentricity is reached, GWs quickly lead to coalescence with the SMBH, in several cases before the orbits of the retrograde BHs can become much more circular. ROs circularize faster in our smaller SMBH example, in two cases reaching circular orbits before merging with the SMBH, despite the GW circularization rate depending more strongly on SMBH mass than the eccentricity driving. They circularize faster because GW circularization is proportional to $1/a^4$ whereas the eccentricity driving is proportional to $a$, and the radiation-pressure-dominated region is smaller for smaller mass SMBHs (see \S\ref{sec:results}). More massive ROs also circularize faster and may reach circular orbits before coalescence. For example a 50~M$_{\rm{\sun}}$ RO with $e_{\rm{0}}$=0.1, will circularize before merging with the SMBH. 

The collision rates per orbit between ROs and prograde orbiting 20~M$_{\rm{\sun}}$ BBHs in the migration trap of a \citet{Sirko:2003aa} AGN disk with a $10^8$~M$_{\sun}$ SMBH are small. \citet{Tagawa_2020} found that BBHs in AGN disks with $10^{7-8}$~M$_{\sun}$ SMBHs dominate the BBH merger rates for the AGN channel. Therefore, the low collision rates found for a $10^8$~M$_{\sun}$ SMBH suggest that ROs will not have a large impact on overall merger rates of low total mass BBHs in AGN disks.

However, \citet{secunda_2020} found that BBHs near the migration trap often grow as massive as 100~M$_{\rm{\sun}}$, and occasionally even 1000~M$_{\rm{\sun}}$. The former BBH mass would increase the probability of interaction for ROs with $e_{\rm{0}}$=0.1 to about $4.3$\%. A 1000~M$_{\rm{\sun}}$ BBH would be almost certain to collide with a RO with $e_{\rm{0}}$=0.1, and has a probability of interaction of $\sim$11\% with a RO with $e_{\rm{0}}$=0.5. However, in our fiducial examples ROs take under a megayear to merge with the SMBH, and in \citep{secunda_2020} these 1000~M$_{\rm{\sun}}$ BBHs take several megayears to form. ROs from further out in the disk or that are ground down from inclined orbits into the disk could perhaps replenish the supply of ROs at later times, although \cite{rauch_1995} and \cite{MacLeod_2020} find that most ROs on inclined orbits will flip to prograde orbits as they align with the disk.

If a RO were to interact with a PO the two could potentially merge or form a BBH. This interaction outcome would be most likely to occur at apocenter of the retrograde BH's orbit where the relative velocities of the POs and ROs would be smallest. A merger would also be more likely if the orbiters are very far out from the central SMBH, where both of their orbital velocities will be lower. However, due to the high relative velocities of POs and ROs, the total interaction energy is likely to be positive, and ROs would most likely act to ionize existing prograde BBHs \citep[e.g.][]{leigh2016,leigh}. For example, the hard-soft boundary describes the binary separation at which a BBH will tend to be disrupted or ionized when it encounters a tertiary. The hard-soft boundary for a RO with $e_{\rm{0}}$=0.1 interacting with a 100~M$_{\rm{\sun}}$ BBH orbiting a $10^8$~M$_{\sun}$ SMBH would be at most \num{5.6e-4}~AU, depending on when the RO and BBH interact. A BBH this compact would likely merge rapidly due to GW emission and not survive long enough to undergo a collision. For comparison, the semi-major axis of a 100~M$_{\rm{\sun}}$ BBH in a migration trap as calculated in eq. \ref{eq:rhill} in \S\ref{sec:prob} would be 4.6~AU.

RO-prograde BBH collisions are far more likely in a \citet{Sirko:2003aa} disk with a $10^6$~M$_{\sun}$ SMBH, where we find collision rates are roughly 8 orders of magnitude higher than in the $10^8$~M$_{\sun}$ case. For the inner radiation-pressure-dominated region of a \citet{Sirko:2003aa} disk this increase is expected as $\tau_{\rm coll}^{-1}$ is roughly proportional to $M^{-4}$ primarily due to the decreased volume of the inner radiation-pressure-dominated region (see \S\ref{sec:results}). While disks around $10^6$~M$_{\sun}$ SMBHs are not expected to be the dominant source of BBH mergers in the AGN merger channel, these rates have implications for whether or not ROs will end up as EMRIs detectable by LISA, which is most sensitive to $10^5 - 10^6$~M$_{\rm{\sun}}$ SMBHs \citep{Babak_2017}. The right panel of Figure~\ref{fig:ret_evolve} suggests that ROs will coalesce with $10^6$~M$_{\rm{\sun}}$ SMBHs, and could still be on eccentric orbits when they merge. Eccentric EMRIs will produce exotic waveforms that would identify them as ROs and may even allow for measurement of gas effects \citep{Derdzinski_2019,Derdzinski_2020}. LISA could even potentially localize its detections to only a few candidate AGN \citep{Babak_2017}. However, if POs are present in the inner disk, the lower panel of Figure \ref{fig:coll_rate} shows ROs will likely collide with them, and these collisions could affect the fate of ROs. Future work is needed to understand how these collisions will impact the EMRI rate and whether a relation between the EMRI rate and the volume of the inner radiation-pressure-dominated disk could help improve our knowledge of AGN disk structure.

Finally, whether ROs will form BBHs with each other is uncertain. ROs' large eccentricities may lead to large relative velocities among them, preventing them from becoming bound. However, if ROs after experiencing orbital decay did undergo a GW inspiral in the innermost disk, they would have a higher probability of being gravitationally lensed by the SMBH, which could be detected by LISA \citep{amaro_2017,nakamura_1998,takahashi_2003,kocsis_2013,dorazio_2019,Chen_2019}. This population of orbiters in the innermost disk could also perturb the inner disk, which may be detectable by electromagnetic observations \citep{McKernan_2013,mckernan14,Blanchard_2017,Ross_2018,Ricci_2020}. 

Here we provide the formalism for calculating the evolution of ROs in an AGN disk and the collision rates of these orbiters with prograde orbiting BBHs. We employ this formalism on a \citet{Sirko:2003aa} model with two example SMBH masses chosen based on importance to the AGN merger channel ($10^{8}$~M$_{\sun}$) and EMRI detectability ($10^{6}$~M$_{\sun}$). Our results show that ROs regularly produce EMRIs. However, the high collision rates of ROs and POs in lower mass SMBH disks could have an impact on EMRIs most easily detected by LISA. On the other hand, the lower collision rates for ROs and POs orbiting higher mass SMBHs suggest that ROs will not have a large effect on the merger rates of the AGN channel. A follow up paper will include a wider parameter study looking at initially inclined orbits,\citeg{Just_2012,Kennedy_2016,Panamarev_2018,MacLeod_2020,FAbj_2020}, higher mass BBHs, varying disk density and scale height profiles, and orbiters beyond the inner radiation-pressure-dominated regime. 

\acknowledgments
We thank the anonymous referee for their thoughtful feedback that improved the quality and clarity of this paper. A.S. would like to thank Charles Emmett Maher for useful conversations. A.S. is supported by a fellowship from the Helen Gurley Brown Revocable Trust and the NSF Graduate Research Fellowship Program under Grant No. DGE-1656466.  NWCL gratefully acknowledges the support of a Fondecyt Iniciacion grant \#11180005. KESF \& BM are supported by NSF AST-1831415 and Simons Foundation Grant 533845. The Flatiron Institute is supported by the Simons Foundation.

\bibliography{bh_bib.bib}{}
\bibliographystyle{aasjournal}

\appendix
\section{Circular Retrograde Orbiters}
\label{sec:circular}
For the special case of a BH on a circular retrograde orbit, $\bm{v} = - \bm{v}_{\rm{disk}}$, where $\bm{v}_{\rm{disk}}$ is the velocity of the disk ($\sqrt{GM/r}$), and the relative velocity between the orbiter and the disk is $\bm{v}_{\rm{rel}}$ = 2$\bm{v}_{\rm{disk}}$. Eqs. \ref{eq:lna_lnE} and \ref{eq:dedt} from \S \ref{sec:evolve} can be used to calculate the evolution of the semimajor axis for a 10~M$_{\rm{\sun}}$ BH on this circular, retrograde orbit around a $10^8$~M$_{\rm{\sun}}$ SMBH in a \citet{Sirko:2003aa} AGN disk. If the BH is initially at a radius of $\sim 10^3 R_{\rm{s}}$, $\Lambda \sim (4M/m)(h/r) \sim 10^5$ and $\rho \sim 10^{-7}$~g~cm$^{-3}$, which gives
\begin{equation}
    \frac{d\ln{a}}{dt} \approx -\frac{1}{{1.5\times10^4}\rm{~yr}}\left(\frac{a}{10^3~R_{\rm{s}}}\right)^3.
\end{equation}

\section{Detailed Derivation of the Collision Rate}
\label{sec:appendix}
Here, we provide more detail to our derivation of the collision rate of a RO (body 1) and a prograde BBH (body 2), which is outlined in \S\ref{sec:prob}. First, we show the derivation of eq.~\eqref{eq:h_bh} for the BH scale height, $h_{\rm{BH}}$.

The eddy turnover speed in a disk will be
\begin{equation}
    \label{eq:v_edd}
    v_{\rm{edd}} \simeq \alpha^{1/2}c_{\rm{s}},
\end{equation}
and the turnover time of eddies of size $l_{\rm{edd}}$ is
\begin{equation}
    \label{eq:tau_edd}
    \tau_{\rm{edd}}=\left(\frac{v_{\rm{edd}}}{l_{\rm{edd}}}\right)^{-1}.
\end{equation}
If we limit $\tau_{\rm{edd}}$ to $\tau_{\rm{edd}}\lesssim \Omega^{-1}$, where $\Omega = (GM/r^3)^{1/2} = c_{\rm{s}}/h$ is the orbital frequency, $l_{\rm{edd}}/h \lesssim v_{\rm{edd}}/c_{\rm{s}} = \alpha^{1/2}$. Therefore the eddy mass is
\begin{equation}
    \label{eq:m_edd}
    m_{\rm{edd}} \simeq \alpha^{3/2} \rho h^3 = \frac{1}{2}\alpha^{3/2} \Sigma h^2.
\end{equation}

Assuming equipartition of vertical kinetic energies gives eq.~\ref{eq:h_bh},
\begin{equation}
    h_{\rm{BH}} \simeq h \frac{v_{\rm{edd}}}{c_{\rm{s}}} \left(\frac{m_{\rm{edd}}}{m_{\rm{bh}}}\right)^{1/2} \simeq h \alpha^{1/2} \left(\frac{\alpha^{3/2} \rho h^3}{m_{\rm{bh}}}\right)^{1/2}.
\end{equation}
The underlying idea for the assumption of equipartition is that turbulence obeys something like the fluctuation-dissipation theorem (FDT) for thermodynamic systems. \cite{Phinney_1992} suggested that the small but measurable eccentricities of binary millisecond pulsars in long-period orbits with white-dwarf companions can be understood as equipartition between epicyclic energy of the orbit and the energies of individual “dominant” convective eddies in the red-giant progenitor of the white dwarf. Observational evidence supports this idea as shown in Figure 8 of \cite{lorimer_2008}. However, the FDT does not apply rigorously to turbulence, so the degree of equipartition probably depends upon the nature of the turbulence. \cite{Nelson_2004} studied the interaction of various masses embedded in a magnetorotationally turbulent disk, with intended application to planets migrating in protostellar disks.  They did not study equipartition explicitly, but they did remark that orbital fluctuations were smaller for larger masses.

Next, we give additional detail on our derivation of eq.~\eqref{eq:vrel} for the relative velocity between the RO and prograde orbiting BBH, $v_{12}=(v_{\phi,1}-v_{\phi,2})^2+(v_{r,1}-v_{r,2})^2 +(v_{z,1}-v_{z,2})^2$. Here, the $\phi$ term is $(\hat{L}_1 - \hat{L}_2)/r^2$. For the $r$ term,
\begin{equation}
    \label{eq:vr}
    v_r=\sqrt{2[\hat{E}-\hat{\Phi}(r)-\hat{L}^2/2r^2]},
\end{equation}
as in eq.~\eqref{eq:annulus}. As in \S\ref{sec:prob} we assume that the $z$ components of the velocities, $v_{z,i}$, are negligible. 

Without making any assumptions about the orbit of the prograde BBH,
\begin{equation}
    \label{eq:full_vrel}
    v_{12}^2 = GM\left[\frac{4}{r} - \frac{1}{a_1} - \frac{1}{a_2}+\frac{2}{r^2}\sqrt{a_1a_2(1-e_1^2)(1-e_2^2)} \pm \frac{1}{\sqrt{a_1a_2}r^2}\sqrt{(r_{+,1}-r)(r_{+,2}-r)(r-r_{-,1})(r-r_{-,2})}\right],
\end{equation}
where the sign of the last term depends on the sign for each $v_r$ given $(E_i,L_i)$.

As mentioned in \S\ref{sec:prob}, because the gas tends to act to dampen the eccentricity of POs we take $e_2 \sim 0$. This approximation eliminates the final term. In addition, since $e_2 \sim 0$, $r =a_2$ giving eq.~\eqref{eq:vrel},
\begin{equation}
    v_{12}=\left[GM\left(\frac{3}{a_2} -\frac{1}{a_1}
    +\frac{2}{a_2^{3/2}}\sqrt{a_1(1-e_1^2)}\right)\right]^{1/2}.
\end{equation}

Now that we have derived a form of $v_{12}$ that is independent of $r$, $\phi$, and $z$, we can remove $v_{12}$ and $\sigma(v_{12})$ from the integral in eq.~\eqref{eq:rate},
\begin{equation}
\tau_{\rm coll}^{-1} = \int dV \frac{d\mathbb{P}_1}{dV}\frac{d\mathbb{P}_2}{dV}\,v_{12}\sigma(v_{12}).
\end{equation}
Written out in its entirety,
\begin{equation}
    \label{eq:rate_detailed}
    \tau_{\rm coll}^{-1}=\frac{v_{12}\sigma(v_{12})}{8\pi^5a_1a_2 h_{\rm BH}^2} \int \exp(-z^2/h_{\rm BH}^2) dz \int  d\phi \int \frac{\delta(r-a_2)}{\sqrt{(r_{+,1}-r)(r-r_{-,1})}} dr.
\end{equation}
Evaluating this integral will give eq.~\eqref{eq:simprate},
\begin{multline}
  \tau_{\rm coll}^{-1} =
  \frac{1}{\sqrt{4\pi h_{\rm BH}^2}}\,\times\,
  \frac{1}{2\pi^3 a_1\sqrt{(a_1(1+e_1)-a_2)(a_2-a_1(1-e_1))}}\,\times\, v_{12}\sigma(v_{12})\,,
\end{multline}
with $h_{BH}$ given by eq.~\eqref{eq:h_bh}, $v_{12}$ given by eq.~\eqref{eq:vrel} and $\sigma(v_{12})$ by eq.~\eqref{eq:sigma}.

\end{document}